\documentclass[a4paper,10pt]{article}
\usepackage[utf8]{inputenc}
\usepackage{amsmath}
\usepackage{amssymb}
\usepackage{graphics}
\usepackage{color}
\usepackage{epsfig}

\title{Tension between $e^+e^-\to\eta\pi^-\pi^+$ and $\tau^-\to\eta\pi^-\pi^0\nu_\tau$ data and non-standard interactions}
\author{Saray Arteaga$^{1}$, Ling-Yun Dai$^{2,3}$, Adolfo Guevara$^{4}$, Pablo Roig$^{5}$\\
$^{1}$The University of Kansas, Department of\\ Physics and Astronomy, 1082 Malott Hall,\\ 1251 Wescoe
Hall Drive, Lawrence, KS 66045, U.S.A.\\
$^{2}$School of Physics and Electronics, Hunan University, \\ Changsha 410082, China.\\
$^{3}$Hunan Provincial Key Laboratory of High-Energy \\ Scale Physics and Applications, Hunan University, \\ Changsha 410082, China.\\
$^{4}$ Institute of Theoretical Physics,
Chinese Academy \\of Sciences, Beijing 100190, China\\
$^{5}$Departamento de Física, Centro de Investigación y de \\Estudios Avanzados del Instituto Politécnico Nacional, \\Apdo. Postal 14-740, 07000 Ciudad de México, México.\\}

\begin{document}

\maketitle

\begin{abstract}
We show the discrepancy between the isospin-rotated $e^+e^-\to\eta\pi^-\pi^+$ cross-section --measured by various collaborations-- and the Belle $\tau^-\to\eta\pi^-\pi^0\nu_\tau$ spectrum, which  cannot be explained by heavy new physics non-standard interactions. We give for the first time the framework needed to study these beyond the standard model  contributions in three-meson tau decays.
\end{abstract}

\section{Introduction}\label{sec:Intro}
In the very accurate isospin symmetry limit, the $e^+e^-\to$ hadrons cross-section is related to the spectral function of semileptonic tau decays (see e.g., ref.~\cite{Eidelman:1990pb}). Beyond tests of this property, based on the conservation of the vector current (CVC), it gave rise to tau-based evaluations of the leading-order  hadronic vacuum polarization ($HVP,LO$) contributions to  the muon g-2 \cite{Alemany:1997tn, Cirigliano:2001er, Cirigliano:2002pv,Davier:2010fmf,Davier:2010nc,Davier:2013sfa,Miranda:2020wdg,Benayoun:2021ody}, $a_\mu=(g_\mu-2)/2$. However,  uncertainties associated to isospin-breaking effects relating both observables are currently too large to make this determination competitive with the one using hadronic $e^+e^-$ cross-section data \cite{Aoyama:2020ynm,Davier:2017zfy,Keshavarzi:2018mgv,Colangelo:2018mtw,Hoferichter:2019mqg,Davier:2019can,Keshavarzi:2019abf,Colangelo:2022jxc}. Notwithstanding, checking the consistency between $\sigma(e^+e^-\to$ hadrons$)$ and exclusive hadron tau decay data is still motivated by the tension exhibited by lattice QCD evaluations \cite{Borsanyi:2020mff,Ce:2022kxy,Alexandrou:2022amy} of $a_\mu^{HVP,LO}$ and the $e^+e^-$-based data-driven extraction \cite{Aoyama:2020ynm}. Depending on which number is  compared to the experimental average of the recent FNAL measurement \cite{Muong-2:2021ojo} and the final result from the BNL experiment \cite{Muong-2:2006rrc}, new physics significance varies sizeably, between barely one and slightly more than four standard deviations. 

In this work, we study the discrepancy -which goes beyond isospin breaking effects-  between both sets of data ($e^+e^-$ and $\tau$) for the exclusive $\eta\pi\pi$ channels, and show that it cannot be explained by heavy new physics.

The isovector component ($I=1$) of the $e^+e^-\to\eta\pi^+\pi^-$ cross section data can be converted to the decay distribution in $\tau^-\to\eta\pi^-\pi^0\nu_\tau$ decays using the approximate conservation of the vector current (CVC), which becomes exact in the isospin symmetry limit \cite{Eidelman:1990pb,Tsai:1971vv,  Cherepanov:2011zz, Dumm:2012vb, Roig:2013ts}:
\begin{eqnarray}\label{CVC}
 \frac{d\Gamma(\tau^-\to\eta\pi^-\pi^0\nu_\tau)}{dQ^2}&=&f(Q^2)\sigma(e^+e^-\to\eta\pi^+\pi^-)|_{I=1}(Q^2)\,,\\
 f(Q^2)&=&\frac{G_F^2\Big|\widetilde{V_{ud}}\Big|^2S_{EW}}{(2\pi)^5M_\tau}\frac{\pi}{4\alpha^2}\left(\frac{M_\tau^2}{Q^2}-1\right)^2 \left(1+2\frac{Q^2}{M_\tau^2}\right)Q^6\,,\nonumber
\end{eqnarray}
with $Q^2=m_{\eta\pi\pi}^2$ the invariant mass squared of the $\eta\pi\pi$ system and $S_{EW}=1.0201(3)$ \cite{SEW} the short-distance electroweak radiative correction. 
We note that $\widetilde{V_{ud}}$ differs from $V_{ud}$ by possible non-standard effects (see  section \ref{sec:EFTNSIs}).

Using eq.~(\ref{CVC}), Belle \cite{Inami:2008ar} data on $\tau^-\to\eta\pi^-\pi^0\nu_\tau$ decays are seen to be incompatible with $e^+e^-\to\eta\pi^-\pi^+$ measurements published by DM2 \cite{Antonelli:1988fw}, ND \cite{Dolinsky:1991vq}, CMD2 \cite{Akhmetshin:2000wv}, BaBar \cite{Aubert:2007ef}, SND \cite{Aulchenko:2014vkn, ACHASOV:2014nra, Achasov:2017kqm} and CMD3 \cite{Gribanov:2019qgw}. We use the best fits obtained in Refs. \cite{Dai:2013joa, Qin:2020udp} to the $e^+e^-\to(\eta/\pi^0)\pi^+\pi^-$ data for the Standard Model prediction. Since possible heavy new physics effects are negligible compared to the photon exchange driving these processes, a possible discrepancy between the isospin-rotated $e^+e^-\to\eta\pi^+\pi^-$ cross-section and the $\tau^-\to\eta\pi^-\pi^0\nu_\tau$ decay rate \cite{Eidelman:1990pb,Tsai:1971vv,  Cherepanov:2011zz, Dumm:2012vb} (besides small isospin breaking effects) could in principle be due to non-standard interactions (NSI) modifying the latter. Thanks to the limits set on possible NSI in semileptonic tau decays \cite{Garces:2017jpz, Miranda:2018cpf, Cirigliano:2018dyk, Rendon:2019awg, Gonzalez-Solis:2019lze, Gonzalez-Solis:2019owk,Arroyo-Urena:2021nil, Arroyo-Urena:2021dfe, Cirigliano:2021yto} we will show that this seeming CVC violation is incompatible with other hadron tau decays data.  Belle-II will improve the measurement of this tau decay channel \cite{Kou:2018nap}, as understanding semileptonic tau decays with eta mesons is required to search for second-class currents and heavy new physics through the discovery of the $\tau^-\to\pi^-\eta^{(\prime)}\nu_\tau$ decays \cite{ Garces:2017jpz,Descotes-Genon:2014tla,Escribano:2016ntp}.\\
The rest of the paper is structured as follows: in section \ref{sec:EFTNSIs} we briefly recall the formalism encoding non-standard interactions in semileptonic tau decays. In section \ref{sec:Amplitude} we derive the $\tau^-\to\eta\pi^-\pi^0\nu_\tau$ decays amplitude, in the Standard Model and including the NSI (involving new hadron contributions, which we account for). In section \ref{sec:NSIeffects} we study the possible effects of NSI on the observables of interest, and show that the discrepancy between $e^+e^-$ and $\tau$ data cannot be explained by heavy new physics, according to the NSI bounds. We conclude in section \ref{sec:Concl}. The appendix summarizes  the setting in which structure-dependent contributions were evaluated.

\section{Effective field theory analysis of NSI in semileptonic tau decays}\label{sec:EFTNSIs}
We consider the most general effective field theory description of $\tau^-\to\bar{u}d\nu_\tau$ decays~\footnote{Hadronic interactions will be considered in the next section.}, assuming massless purely left-handed neutrinos \cite{Garces:2017jpz, Miranda:2018cpf, Cirigliano:2018dyk, Rendon:2019awg, Gonzalez-Solis:2019lze, Gonzalez-Solis:2019owk,Arroyo-Urena:2021nil, Arroyo-Urena:2021dfe, Cirigliano:2021yto} (see e.~g.~ refs. \cite{Cirigliano:2009wk, Bhattacharya:2011qm, Cirigliano:2012ab, Cirigliano:2013xha, Chang:2014iba, Courtoy:2015haa,Gonzalez-Alonso:2016etj, Gonzalez-Alonso:2017iyc, Alioli:2017ces, Gonzalez-Alonso:2018omy, Descotes-Genon:2018foz} for other semileptonic processes involving light quarks within this framework), which only rests on the $SU(3)_C\times U(1)$ local gauge symmetry below the electroweak scale. For later convenience we introduce $\epsilon_V:=\epsilon_L+\epsilon_R$ and $\epsilon_A:=-\epsilon_L+\epsilon_R$, so that the relevant Lagragian at dimension six is
\begin{eqnarray}
 \mathcal{L}&=&-\frac{G_F \widetilde{V_{ud}}}{\sqrt{2}}\Big\lbrace\bar{\tau}\gamma^\mu(1-\gamma_5)\nu_\tau\cdot\Big[\bar{u}\gamma_\mu(1-\gamma_5)d+\bar{u}\gamma_\mu(\epsilon_V^\tau+\epsilon_A^\tau \gamma_5)d\Big]\\
 &+&\bar{\tau}(1-\gamma_5)\nu_\tau\cdot \bar{u}(\epsilon_S^\tau-\epsilon_P^\tau\gamma_5)d+2\epsilon_T^\tau\bar{\tau}\sigma^{\mu\nu}(1-\gamma_5)\nu_\tau\cdot \bar{u}\sigma_{\mu\nu}d\Big\rbrace+\mathrm{h.c.},\nonumber
\end{eqnarray}
where $G_F V_{ud}=G_F \widetilde{V_{ud}}(1+\epsilon_V^e)$ \cite{Descotes-Genon:2018foz}. We neglect higher-dimensional operators, suppressed by powers of $M_\tau/\Lambda$, since current limits on the $\epsilon_i$ coefficients \cite{Garces:2017jpz, Miranda:2018cpf, Cirigliano:2018dyk, Rendon:2019awg, Gonzalez-Solis:2019lze, Gonzalez-Solis:2019owk,Arroyo-Urena:2021nil, Arroyo-Urena:2021dfe, Cirigliano:2021yto} correspond to $\Lambda\sim\mathcal{O}(\mathrm{TeV})$ (under the weak-coupling hypothesis). As we only compute CP-conserving observables~\footnote{See e.g., refs. \cite{Rendon:2019awg,Cirigliano:2017tqn, Chen:2019vbr,Chen:2020uxi,Chen:2021udz} for studies of CP violation in $\tau\to K_S\pi\nu_\tau$ decays within this low-energy effective field theory.}, the $\epsilon_i$ coefficients are taken real. They are translated straightforwardly \cite{Cirigliano:2009wk, Gonzalez-Alonso:2017iyc} into the SMEFT \cite{Buchmuller:1985jz, Grzadkowski:2010es} couplings. For vanishing $\epsilon_i$, the SM is recovered. We will work in the $\overline{MS}$ scheme, at a scale $\mu=2$ GeV.

\section{$\tau^-\to\eta\pi^-\pi^0\nu_\tau$ amplitude}\label{sec:Amplitude}
We will assign momenta as~\footnote{Unlike ref. \cite{Kumar:1970cr}, we do not use $Q$ as the momentum of the decaying particle, since this corresponds to the invariant mass of the $\eta\pi\pi$ system in our notation.} $\tau^-(P)\to\nu_\tau(p_1)\eta(p_2)\pi^-(p_3)\pi^0(p_4)$ and use Kumar kinematics \cite{Kumar:1970cr}~\footnote{In ref.\cite{Saray} it was shown that the kinematics adopted in e. g., ref. \cite{Dumm:2012vb} is not appropriate when tensor interactions are considered, as the factorization of the lepton and hadron parts (with the latter only depending on three independent invariants which can be written in terms of the meson momenta) no longer holds.}, so that the outermost integration variable, $(P-p_1)^2=(p_2+p_3+p_4)^2$, gives us $Q^2=m_{\eta\pi\pi}^2$, whose distribution was measured by Belle \cite{Inami:2008ar}.

\subsection{Hadronization: Standard Model and beyond}\label{sec:Hadronization}
In the Standard Model, the $\tau^-\to\eta\pi^-\pi^0\nu_\tau$ decay amplitude is
\begin{equation}\label{eq:decay_amp}
{\cal M}  \, = \,  - \, \frac{G_F}{\sqrt{2}} \, %\widetilde{
V_{ud}%}
\, \sqrt{S_{EW}}\, \bar u_{\nu_\tau} \gamma^\mu\, (1-\gamma_5) u_\tau\, \mathcal{H}_\mu \; ,
\end{equation}
where $\mathcal{H}_\mu$ encodes the hadronization into the three final-state mesons ($h_1=\eta,h_2=\pi^-,h_3=\pi^0$  in our case and with our conventions). Lorentz invariance determines the most general decomposition of $\mathcal{H}^\mu$ to be
\begin{eqnarray} \label{generaldecomposition_3mesons}
\mathcal{H}^\mu & = & \left\langle h_1(p_2)h_2(p_3)h_3(p_4)|(V-A)^\mu|0\right\rangle = F_1^A(Q^2,\,s_2,\,t_3')\, V_1^\mu\, \\
& & \hspace{-1.0cm} + \; F_2^A(Q^2,\,s_2,\,t_3')\,V_2^\mu\;
+\; i\, F_3^V(Q^2,\,s_2,\,t_3')\,V_3^\mu\; +
\; F_4^A(Q^2,\,s_2,\,t_3')\, Q^\mu \, , \nonumber
\end{eqnarray}
where the chosen set of independent Lorentz structures is
\begin{eqnarray} \label{Set_of_independent_Vectors_3meson}
& & V_1^\mu\, = \, \left( g^{\mu\nu} - \frac{Q^{\mu}Q^{\nu}}{Q^2}\right) \,
(p_2 - p_4)_{\nu} \,\,\, , \quad V_2^\mu\, = \, \left( g^{\mu\nu} -
\frac{Q^{\mu}Q^{\nu}}{Q^2}\right) \,
(p_3 - p_4)_{\nu} \, ,\nonumber\\
& & V_3^\mu\, =
\,\varepsilon_{\mu\alpha\beta\gamma}\,p_2^\alpha\,p_3^\beta\,p_4^\gamma,\hspace*{17ex}
Q^\mu\,=\,(p_2\,+\,p_3\,+\,p_4)^\mu\,,\nonumber\\ 
& & s_2\,=\,(Q\,-\,p_2)^2\ ,\hspace*{22ex} t_3'\,=\,(p_2+p_4)^2\nonumber\\
\end{eqnarray}
and the relevant form factors ($F_i, i=1,...,4$) are driven by either vector or axial-vector currents (as indicated by their superscript, $V/A$) and carry quantum numbers of pseudoscalar ($F_4$), vector ($F_3$) or axial-vector ($F_{1,2}$) degrees of freedom. Very approximate $G$-parity conservation by the strong interactions ~\footnote{$G$-parity is built from $C$-parity and isospin symmetry. For consistency, ignoring the effect of the $F_{1,2,4}$ form factors requires to describe $F_3$ in the
isospin symmetry limit.} produces vanishing axial-vector form factors in this channel, in such a way that -to an excellent accuracy- the dynamics of the considered decays are driven solely by the vector form factor, $F_3$%, which is related to the $W_B$ structure function appearing in eq.~(\ref{Q2 spectrum}) by $W_B(Q^2,s_1,s_2)=V_3^2\,|F_3^V(Q^2,s_1,s_2)|^2$. It can be easily checked that%~\footnote{We note that $Q^2=s_1+s_2+s_3-2m_\pi^2-m_\eta^2$.}
%\begin{eqnarray}
%V_3^2 & = & m_\eta^2 m_\pi^4-m_\eta^2 \left(\frac{s_1-2 m_\pi^2}{2}\right)^2-m_\pi^2 \left(\frac{s_3-m_\pi^2-m_\eta^2}{2}\right)^2- \\
%& & m_\pi^2 \left(\frac{s_2-m_\pi^2-m_\eta^2}{2}\right)^2 + \frac{(s_3-m_\pi^2-m_\eta^2)(s_2-m_\pi^2-m_\eta^2)}{2} \frac{s_1-2 m_\pi^2}{2}\,.\;\nonumber
%\end{eqnarray}
, which will be taken from the best fits of Refs. \cite{Dai:2013joa, Qin:2020udp}.

As explained, $F_{1,2,4}^A$ vanish in the limit of $G$-parity conservation. We will however compute the isospin-breaking contributions to these form factors given by scalar resonance  exchanges. Our motivation to include these subleading effects only for the scalar mesons contributions is two-folded: on the one hand isospin-violating $f_0-a_0$ mixing is enhanced with respect to other isospin breaking effects by the approximate degeneracy of these states and their comparable value to the kaon-antikaon thresholds \cite{Achasov:1979xc}. On the other hand, Belle-II shall measure the di-meson mass spectra \footnote{Unexpectedly, Belle \cite{Inami:2008ar} found disagreement between their measured $\pi^-\pi^0$ spectra and the Monte Carlo event generator \cite{Jadach:1993hs, Actis:2010gg}, validated with precise previous data on the weak pion vector form factor.} in $\tau^-\to\eta\pi^-\pi^0\nu_\tau$ decays and a theoretically-motivated parametrization of scalar meson exchanges in these processes will benefit their analysis. In this way, we will construct the hadronic input needed for NSI contributions to the considered decays.

There are three possible contributions with intermediate scalar resonances (all of them in the axial-vector current), one per channel. Schematically, they are:\\
- $A^\mu\to a_0^- \eta$, with $a_0^-\to\pi^-\pi^0$ via $\eta^{(\prime)}-\pi^0$ mixing, in the $s_1$ channel ($T^\alpha_{s_1}$ below).\\
- $A^\mu\to a_0^0 \pi^-$, with $a_0^0\to\pi^0\eta^{(\prime)}$ via $f_0-a_0$ mixing, in the $s_2$ channel ($T^\alpha_{s_2}$ below).\\
- $A^\mu\to a_0^- \pi^0$, with $a_0^-\to\pi^-\eta^{(\prime)}$ via $\eta^{(\prime)}-\pi^0$ mixing, in the $s_3$ channel ($T^\alpha_{s_3}$ below).\\

%It is thus evident that scalar resonances will not contribute to the (vector current mediated) $e^+e^-\to\eta\pi^+\pi^-$ but only (very mildly) to $\tau^-\to\eta\pi^-\pi^0\nu_\tau$ decays.
%Therefore, one should fix the vector current parameters either using only low-energy $e^+e^-$ data or also $\tau^-$ decay data and, \textit{a posteriori} examine the possible tiny effect of scalar resonance exchange in $\tau^-\to\eta\pi^-\pi^0\nu_\tau$ decay data.\\

The corresponding scalar resonance exchange contributions, computed within Resonance Chiral Theory \cite{RChT} (R$\chi$T, see appendix), are 
\begin{eqnarray}\label{Scalarcontributions}
T_{s_1}^\alpha&=&p_2^\alpha\,\frac{4\sqrt{2}C_{q^{(\prime)}}}{F^3}\frac{C_q \epsilon_{\pi\eta}+C_{q'} \epsilon_{\pi\eta'}}{M_{a_0}^2-s_1-iM_{a_0}\Gamma_{a_0}(s_1)}
c_d\left[c_d\left(\frac{s_1}{2}-m_\pi^2\right)+c_m m_\pi^2\right]\,,\\
T_{s_2}^\alpha&=&p_3^\alpha\,\frac{4\sqrt{2}C_{q^{(\prime)}}}{\sqrt{3}F^3}\frac{\epsilon_{a_0f_0}(s_2)}{M_{a_0}^2-s_2-iM_{a_0}\Gamma_{a_0}(s_2)}
c_d\left[c_d\left(\frac{s_2-m_\pi^2-m_\eta^2}{2}\right)+c_m m_\pi^2\right]\,,\nonumber\\
T_{s_3}^\alpha&=&p_4^\alpha\,\frac{4\sqrt{2}C_{q^{(\prime)}}}{F^3}\frac{C_q \epsilon_{\pi\eta}+C_{q'} \epsilon_{\pi\eta'}}{M_{a_0}^2-s_3-iM_{a_0}\Gamma_{a_0}(s_3)}
c_d\left[c_d\left(\frac{s_3-m_\pi^2-m_\eta^2}{2}\right)+c_m m_\pi^2\right]\,,\nonumber
\end{eqnarray}
where~\cite{Hanhart:2007bd} $\epsilon_{a_0f_0}(s_2)=\epsilon_{a_0f_0}(\sigma_{K^0}(s_2)-\sigma_{K^+}(s_2))/2$, with $\epsilon_{a_0f_0}\sim\mathcal{O}(1)$~\footnote{$T_{s_2}^\alpha$ has been obtained assuming, for simplicity, that $f_0(980)$ is a pure octet state. If it comes from the mixing of the octet and singlet $f$ states, the corresponding mixing coefficient can be absorbed in the constant $\epsilon_{a_0f_0}$, that we will fix to unity for definiteness. This and other ambiguities present in the description of the scalar mesons (like possible tetraquark components \cite{Zyla:2020zbs} and more complicated mixing pattern \cite{Cirigliano:2003yq}) prevent us from attempting to derive the real part of the meson-meson loops which should be present in the $a_0$ propagators in eq.~(\ref{Scalarcontributions}) to fulfill analyticity.}. Short-distance QCD constraints set $4 c_d c_m = F^2$ \cite{Jamin:2000wn, Jamin:2001zq} (with $F\sim 92$ MeV) and $c_m \sim 3 c_d$ is preferred phenomenologically (see \cite{Escribano:2010wt} and references therein).
$C_{q^{(\prime)}}$ are given in terms of the $\eta-\eta'$ mixing parameters \cite{DoubleAngleMixing}
\begin{equation}
  C_q  \equiv  \frac{1}{\sqrt{3}\mathrm{cos}(\theta_8-\theta_0)}\left(\frac{\mathrm{cos}\theta_0}{f_8}-\frac{\sqrt{2}\mathrm{sin}\theta_8}{f_0}\right)\,,
  \,
  C_{q^\prime}\equiv\frac{1}{\sqrt{3}\mathrm{cos}(\theta_8-\theta_0)}\left(\frac{\sqrt{2}\mathrm{cos}\theta_8}{f_0}+\frac{\mathrm{sin}\theta_0}{f_8}\right)\,,
\end{equation}
and $\varepsilon_{\pi\eta}=(9.8\pm0.3)\cdot10^{-3}$ and $\varepsilon_{\pi\eta^{\prime}}=(2.5\pm1.5)\cdot10^{-4}$~\cite{Escribano:2016ntp}. We will take the numerical values for $C_{q^{(\prime)}}$ from Ref. \cite{Guevara:2018rhj} (see also ref. \cite{Roig:2014uja}):
$C_q=0.69\pm0.03$ and $C_{q^\prime}=0.60\pm0.03$, obtained in the chiral limit.

The $\Gamma_{a_0}(s_i)$ energy-dependent width is given by \cite{Escribano:2016ntp}
\begin{equation}
\Gamma_{a_0}(s_i)=\Gamma_{a_0}(M_{a_0}^{2})\left(\frac{s_i}{M_{a_0}^{2}}\right)^{3/2}\frac{h(s_i)}{h(M_{a_0}^{2})}\ ,
\end{equation}
with ($\sigma_{PQ}(s_i)=\lambda^{1/2}(s_i,m_P^2,m_Q^2)/s\times\Theta(s_i-(m_P+m_Q)^2)$ is a kinematical factor and $\lambda(a,b,c)=(a-b-c)^2-4bc$)
\begin{equation}
\begin{array}{l}
h(s_i)=
\displaystyle{\sigma_{K^-K^0}(s_i)
+2\cos^2\phi_{\eta\eta^\prime}\left(1+\frac{\Delta_{\pi^{-}\eta}}{s_i}\right)^{2}\sigma_{\pi^{-}\eta}(s_i)}\\[4ex]
\qquad\quad
\displaystyle{
+2\sin^2\phi_{\eta\eta^\prime}\left(1+\frac{\Delta_{\pi^{-}\eta^{\prime}}}{s_i}\right)^{2}\sigma_{\pi^{-}\eta^{\prime}}(s_i)}\ ,
\end{array}
\end{equation}
$\Delta_{PQ}=m_P^2-m_Q^2$ and $\phi_{\eta\eta^\prime}=(41.4\pm 0.5)^\circ$~\cite{Ambrosino:2009sc}. Mass and on-shell width of the $a_0$ resonance will be taken from the PDG \cite{Zyla:2020zbs}. Scalar contributions in eq.~(\ref{Scalarcontributions}) can be written in terms of the $F_{1,2,4}^A$ form factors using
\begin{eqnarray}
 p_2^\mu & = & \frac{Q^\mu}{2Q^2}(Q^2-s_1+m_\eta^2)+\frac{2V_1^\mu}{3}-\frac{V_2^\mu}{3}\,,\\
 p_3^\mu & = & \frac{Q^\mu}{2Q^2}(Q^2-s_2+m_\pi^2)-\frac{V_1^\mu}{3}+\frac{2V_2^\mu}{3}\,,\nonumber\\
 p_4^\mu & = & \frac{Q^\mu}{2Q^2}(s_1+s_2-m_\eta^2-m_\pi^2)-\frac{V_1^\mu}{3}-\frac{V_2^\mu}{3}\,.\nonumber
\end{eqnarray}

For consistency --as scalar resonance contributions are included-- axial-vector current contributions induced from $\tau^-\to\pi^-\pi^0\pi^0$ decays coming from $\pi^0-\eta$ mixing need to be accounted for as well. This is done following references \cite{Dumm:2009va, Shekhovtsova:2012ra, Nugent:2013hxa} (including also the $KK\pi$ cuts \cite{Dumm:2009kj} into the energy-dependent $\Gamma_{a_1}$). The overall factor $\epsilon_{\pi\eta}^2$ suppresses strongly this contribution, which does not introduce any additional free parameter.\\\\
Beyond the Standard Model, the vector and axial-vector matrix elements (corresponding to the $\bar{d}\gamma^\mu u$ and  $\bar{d}\gamma^\mu \gamma_5 u$ quark currents) can be written in terms of the $\left\lbrace F_i\right\rbrace_{i=1,...,4}$ form factors (we omit their dependence on $Q^2,s_1,s_2$ below)
\begin{equation}
 H^\mu_V=iF_3^{V,\text{NSI}} V_3^\mu\,,\quad -H^\mu_A=F_1^{A,\text{NSI}} V_1^\mu+F_2^{A,\text{NSI}} V_2^\mu+F_4^{A,\text{NSI}}Q^\mu\,,
\end{equation}
which are defined by the currents
\begin{equation}
H_{(V/A)}^\mu
= \epsilon_{(V/A)}\langle \eta \pi^- \pi^0| \bar{d}\gamma^\mu(1,\gamma_5) u| 0 \rangle. 
\end{equation}
We can relate the previous terms with the hadronization in eq. (\ref{eq:decay_amp}) with the relation 
\begin{equation}
 \mathcal{H}^\mu= H_\text{NSI}^\mu + H_L^\mu,
\end{equation}
where $H_L$ contains all SM interactions and $H_{V/A}^\mu\in H_\text{NSI}$.
The (pseudo)scalar matrix elements can be related to the former using Dirac equation. This shows the vanishing of the hadron matrix element of the scalar current, while the pseudoscalar one (for the $\bar{d}\gamma_5 u$ quark current) can be related to $H^\mu_A$, 
which is defined as 
\begin{equation}
    H_P=\epsilon_P\langle \eta \pi^- \pi^0| \bar{d}\gamma_5 u| 0 \rangle, 
\end{equation}
yielding
\begin{equation}\label{HP&F4}
 H_P=\frac{F_4^A Q^2}{m_u+m_d}\,.
\end{equation}

We will finally address the hadronization of the tensor current ($\langle \eta \pi^- \pi^0| \bar{d}\sigma^{\mu\nu} u|0\rangle$)  for which we will employ Chiral Perturbation Theory \cite{ChPT} with tensor sources \cite{Cata:2007ns}. The leading contribution in the chiral counting is given in terms of a single coupling constant, $\Lambda_2$, which can be determined from the lattice \cite{Baum:2011rm} to be $\Lambda_2=(11.1\pm0.4)$ MeV \cite{Gonzalez-Solis:2019lze}. In terms of it, the hadron matrix element for the tensor current is
\begin{equation}
 H^{\mu\nu}_T=i\frac{\Lambda_2 C_q}{\sqrt{2}F^3}\epsilon^{\mu\nu\alpha\beta}(p_{3\alpha}p_{2\beta}-p_{2\alpha}p_{3\beta})\,.
\end{equation}

\subsection{Decay amplitude}\label{sec:GeneralizedAmplitude}

The $\tau^-\to\eta\pi^-\pi^0\nu_\tau$ decay amplitude can be written~\footnote{Although the effect of the short-distance radiative electroweak corrections encoded in $S_{EW}$ affects only the SM contribution, we approximate it as a global factor in the equation below. Its accuracy is sufficient for our precision and renders simpler expressions.}
\begin{eqnarray}\label{initialamplitude}
 \mathcal{M}&=&\mathcal{M}_{SM}+\mathcal{M}_V+\mathcal{M}_A+\mathcal{M}_P+\mathcal{M}_T\\
 &=&-\frac{G_F\widetilde{V_{ud}}\sqrt{S_{EW}}}{\sqrt{2}}\Big[L_\mu (H^\mu+\epsilon_V H^\mu_V+ \epsilon_A H^\mu_A)-\epsilon_P LH_P+2\epsilon_TL_{\mu\nu}H^{\mu\nu}\Big]\,,\nonumber
 \end{eqnarray}
where the following lepton currents were introduced
\begin{equation}
 L = \bar{u}(p_1)(1+\gamma_5)u(P)\,,\quad L_\mu = \bar{u}(p_1)\gamma_\mu(1-\gamma_5)u(P)\,,\quad L_{\mu\nu} = \bar{u}(p_1)\sigma_{\mu\nu}(1+\gamma_5)u(P)\,.
\end{equation}
Using Dirac equation, $L_\mu Q^\mu=M_\tau L$ is obtained. This, together with eq. (\ref{HP&F4}), allows the convenient rewriting $\epsilon_A L_\mu H^\mu_A-\epsilon_P L H_P=\epsilon_A L_\mu H'^\mu_A$, where
\begin{equation}
 H'^\mu_A=H^\mu_A-\frac{\epsilon_P}{\epsilon_A}\frac{F_4^AQ^2Q^\mu}{M_\tau(m_u+m_d)}\,,
\end{equation}
which in turn allows to recast eq. (\ref{initialamplitude}) as
\begin{eqnarray}\label{amplitude}
 \mathcal{M}&=&\mathcal{M}_{SM}+\mathcal{M}_V+\mathcal{M}_{A'}+\mathcal{M}_T\\
 &=&-\frac{G_F\widetilde{V_{ud}}\sqrt{S_{EW}}}{\sqrt{2}}\Big[L_\mu (H^\mu+\epsilon_V H^\mu_V+ \epsilon_A H'^\mu_A)+2\epsilon_TL_{\mu\nu}H^{\mu\nu}\Big]\,,\nonumber
 \end{eqnarray}
 that we have used to compute the observables presented in the following section. We provide an ancillary file with the analytic results for the different contributions to $|\mathcal{M}|^2$, for which we used FeynCalc \cite{Shtabovenko:2020gxv,Shtabovenko:2016sxi,Mertig:1990an}.

\section{CVC prediction of the $\tau^-\to\eta\pi^-\pi^0\nu_\tau$ decay rate and NSI}\label{sec:NSIeffects}
\begin{figure}[hpt]
\begin{center}
\includegraphics[width=0.8\textwidth,height=0.35\textheight]{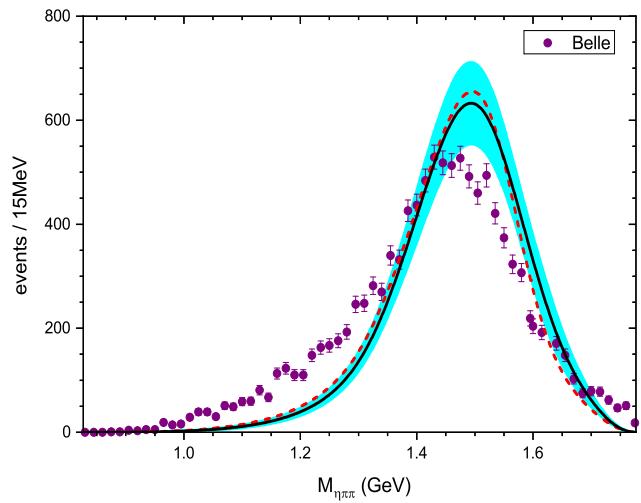}
\caption{\label{Fig:ee2tau} The prediction of $\tau^-\to\eta\pi^-\pi^0\nu_\tau$ from $e^+e^-\to\eta\pi^+\pi^-$ amplitudes. The black solid line is calculated from Fit II of Ref. \cite{Qin:2020udp} and the red dashed line is from Fit 4 of Ref. \cite{Dai:2013joa}. The cyan band describes the uncertainty obtained by a combined statistics of the error band of Fit II in  Ref.~\cite{Qin:2020udp} and the difference between the red and black lines. Belle data are represented by purple dots. }
\end{center}
\end{figure}

For our isospin-rotated prediction of the $\tau^-\to\eta\pi^-\pi^0\nu_\tau$ decays in absence of new physics we  will use the CVC relation (see eq.~(\ref{CVC})), with $e^+e^-\to\eta\pi^+\pi^-$ given by the best fit solutions of refs. \cite{Dai:2013joa,Qin:2020udp} %(see also ref. \cite{Volkov:2013zba})
.
Specifically, Fit 4 in ref. \cite{Dai:2013joa} and Fit II in ref. \cite{Qin:2020udp}, respectively.
The amplitudes were calculated using R$\chi$T  \cite{RChT} and confronted with the latest high statistics experimental measurements of $e^+e^-\to\eta\pi^+\pi^-$ cross sections up to 2.3~GeV, including those of Babar \cite{Aubert:2007ef}, SND \cite{Aulchenko:2014vkn,Achasov:2017kqm},
and CMD3 \cite{Gribanov:2019qgw}.
%%%
By isospin rotation, the prediction of the invariant mass spectrum of $\tau^-\to\eta\pi^-\pi^0\nu_\tau$ decays is given in Fig.\ref{Fig:ee2tau}.
%%%
It can be seen that the prediction from $e^+e^-\to\eta\pi^+\pi^-$ is quite different from that of the Belle data \cite{Inami:2008ar}, especially in the region of 0.9-1.4~GeV.  The $\tau\to\eta\pi^-\pi^0\nu_\tau$ branching ratio   is $(1.71\pm0.13)\times10^{-3}$, using the Fit II in  Ref.~\cite{Qin:2020udp}, and  $(1.55\pm0.18)\times10^{-3}$ from Fit 4 of Ref.~\cite{Dai:2013joa}. The PDG quotes $(1.39\pm0.07)\times10^{-3}$  instead, from which our previous numbers are $2.2$ and $0.8$ $\sigma$ away, respectively. 
Meanwhile, $e^+e^-\to\eta\pi^+\pi^-$ data are considered much more accurate and trustworthy. Hence, it would be rather important for Belle-II to improve the measurement of this decay channel in the future.

The effects of NSI are constrained thanks to %the analysis of ref. \cite{Cirigliano:2018dyk}, where a combination of inclusive and exclusive analysis of semileptonic tau decays yielded (limits on $\epsilon^\tau_S$ are only quoted for completeness)
the most recent determination (in agreement with previous ones) of these couplings \cite{Cirigliano:2021yto}, yielding
\begin{equation}\label{Ciriglianobounds2}
\begin{pmatrix}
\epsilon^\tau_L-\epsilon^e_L+\epsilon^\tau_R-\epsilon^e_R\\
\epsilon^\tau_R\\
%\epsilon^\tau_S\\
\epsilon^\tau_P\\
\epsilon^\tau_T\\
\end{pmatrix}
=
\begin{pmatrix}
 2.4\pm2.6\\
 0.7\pm1.4\\
% -0.6\pm1.5\\
 0.4\pm1.0\\
 -3.3\pm6.0\\
\end{pmatrix}
\cdot 10^{-2}\,,
\end{equation}
with the correlation matrix
\begin{equation}\label{Ciriglianocorr}
 C_\epsilon=\begin{pmatrix}
   1&0.87 & -0.18 & -0.98\\
   & 1& -0.59 & -0.86\\
   &       &  1 & 0.18  \\
   & & & 1 \\
 \end{pmatrix}\,.
\end{equation}

 \begin{figure}[!ht]
 \centering\includegraphics[scale=0.85]{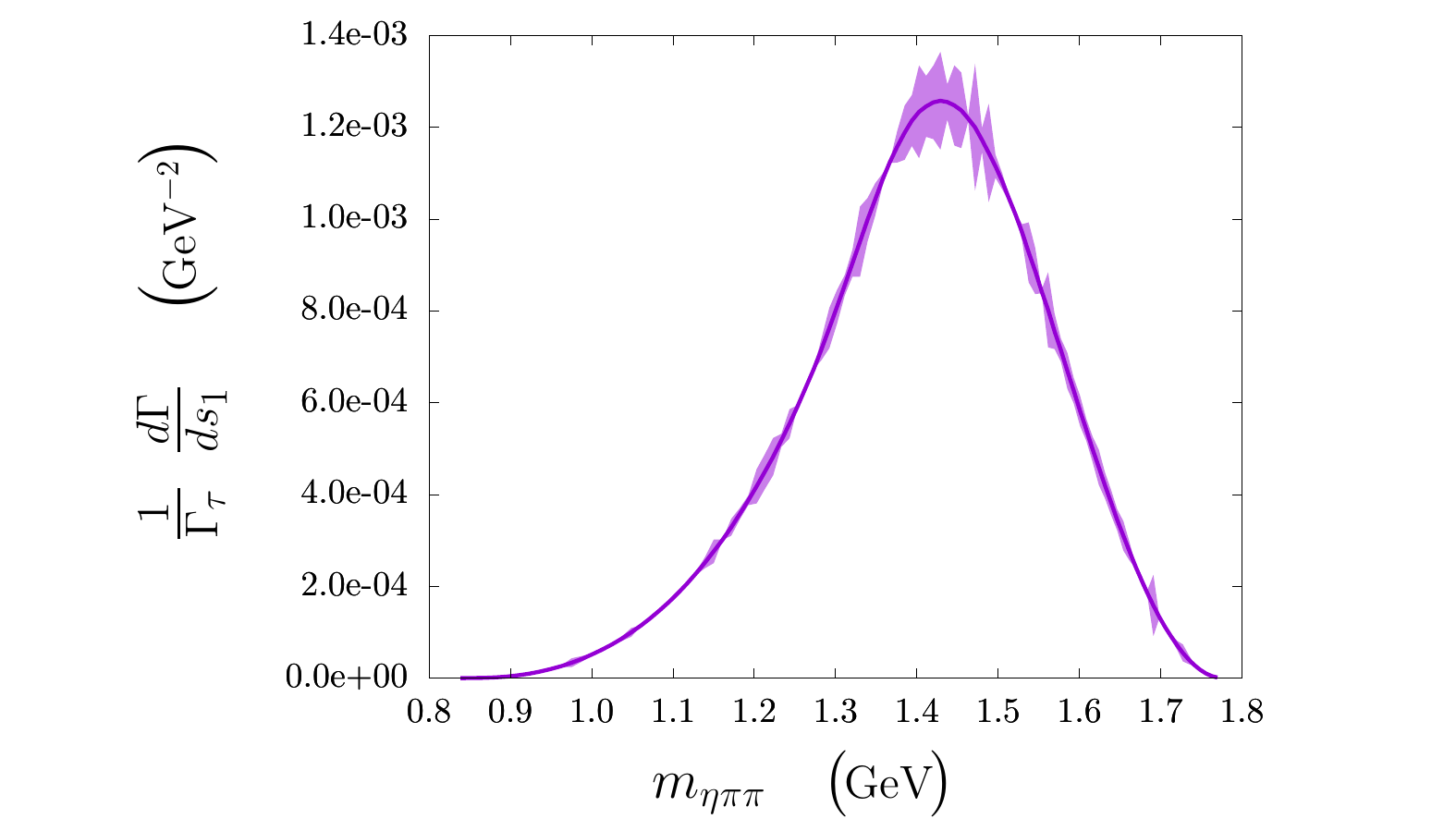}\caption{Invariant mass spectrum of the $\tau^-\to\eta\pi^-\pi^0\nu_\tau$ transition. The purple stripe is the error band obtained from the Gaussian variation of parameters at each bin, the solid purple line represents the mean of the distribution.}\label{fig:InvMassSpectrum}
 \end{figure}

%Using these constrictions on the NSI parameters, we computed the Branching Fraction $\mathfrak{B}$
In our numerical analysis we used 2500 points\footnote{This amount of points was chosen to obtain a kurtosis near to 3, getting $K=3.17$, which guarantees their distribution is Gaussian.} generated randomly following a Gaussian distribution using the parameters and errors in eq. (\ref{Ciriglianobounds2}) and the correlation matrix of eq. (\ref{Ciriglianocorr}). The vector form factor in ref. \cite{Dumm:2012vb} was used  in the following.%, obtaining 
% \begin{equation}
%  \mathfrak{B}(\tau^-\to\eta\pi^-\pi^0\nu_\tau)=(1.60 \pm 0.20\pm0.04)\cdot10^{-3}\,,\label{eq:BRwithNSI}
% \end{equation}
% where the first error comes from the model dependence of our description and the second one from the possible NSI contribution, which is much smaller than the former (we recall that the PDG error on this branching ratio is $0.07\times10^{-3}$).
 
 We also computed the invariant mass $m_{\eta\pi\pi}$ spectrum for which we again  used a Gaussian  variation of the parameters, generating 2500 points at each bin of the spectrum, shown in figure \ref{fig:InvMassSpectrum}. 

 \begin{figure}[!ht]
  \centering\includegraphics[scale=0.85]{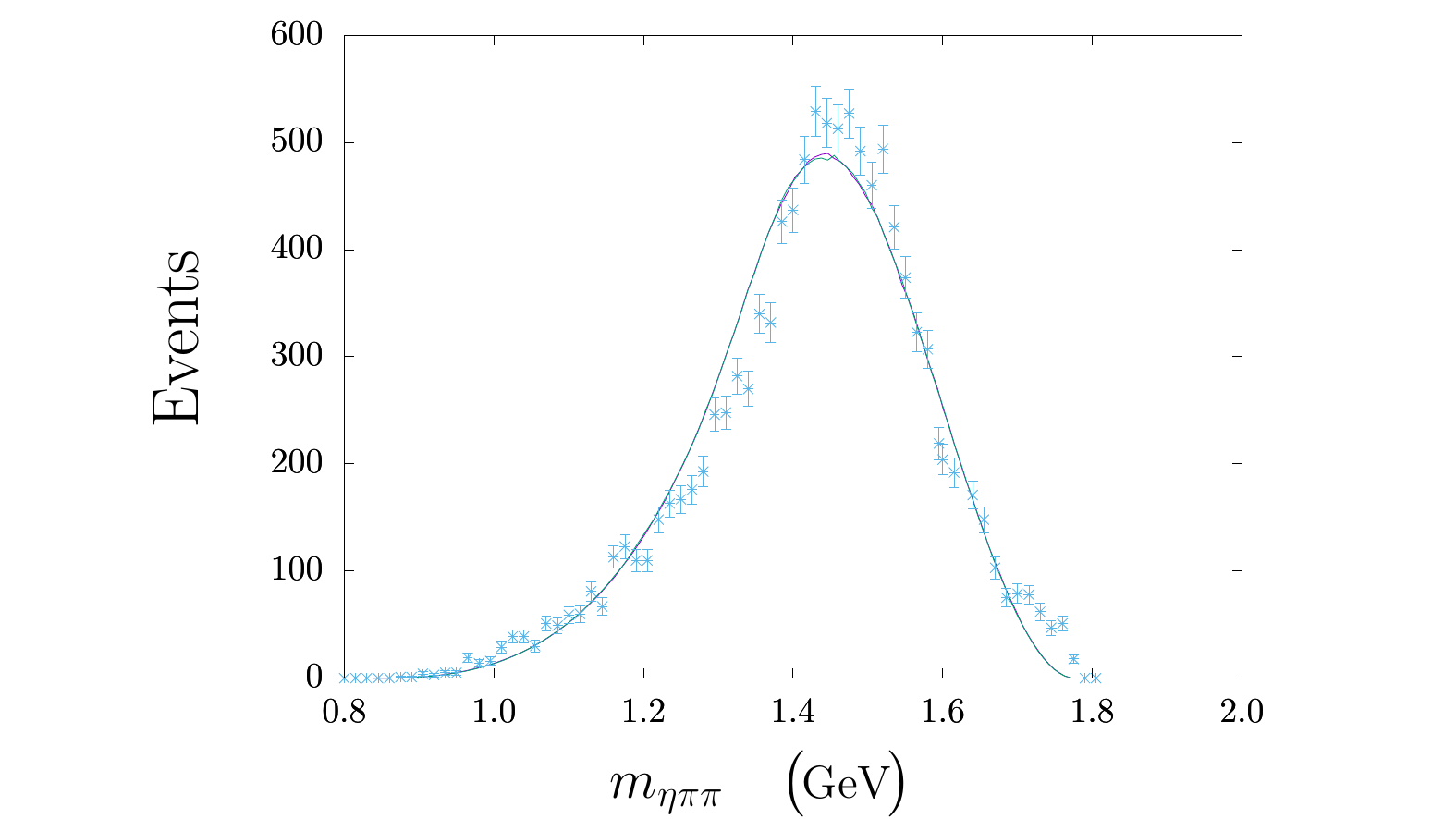}\caption{Comparison between the spectra obtained using the complete amplitude (purple line) with the NSI turned off (green line) and the Belle data (pale blue dots). The difference between both lines can be slightly appreciated only near the peak.}\label{fig:SMvsNSI}
 \end{figure}

\begin{figure}[!ht]
 \centering\includegraphics[scale=0.85]{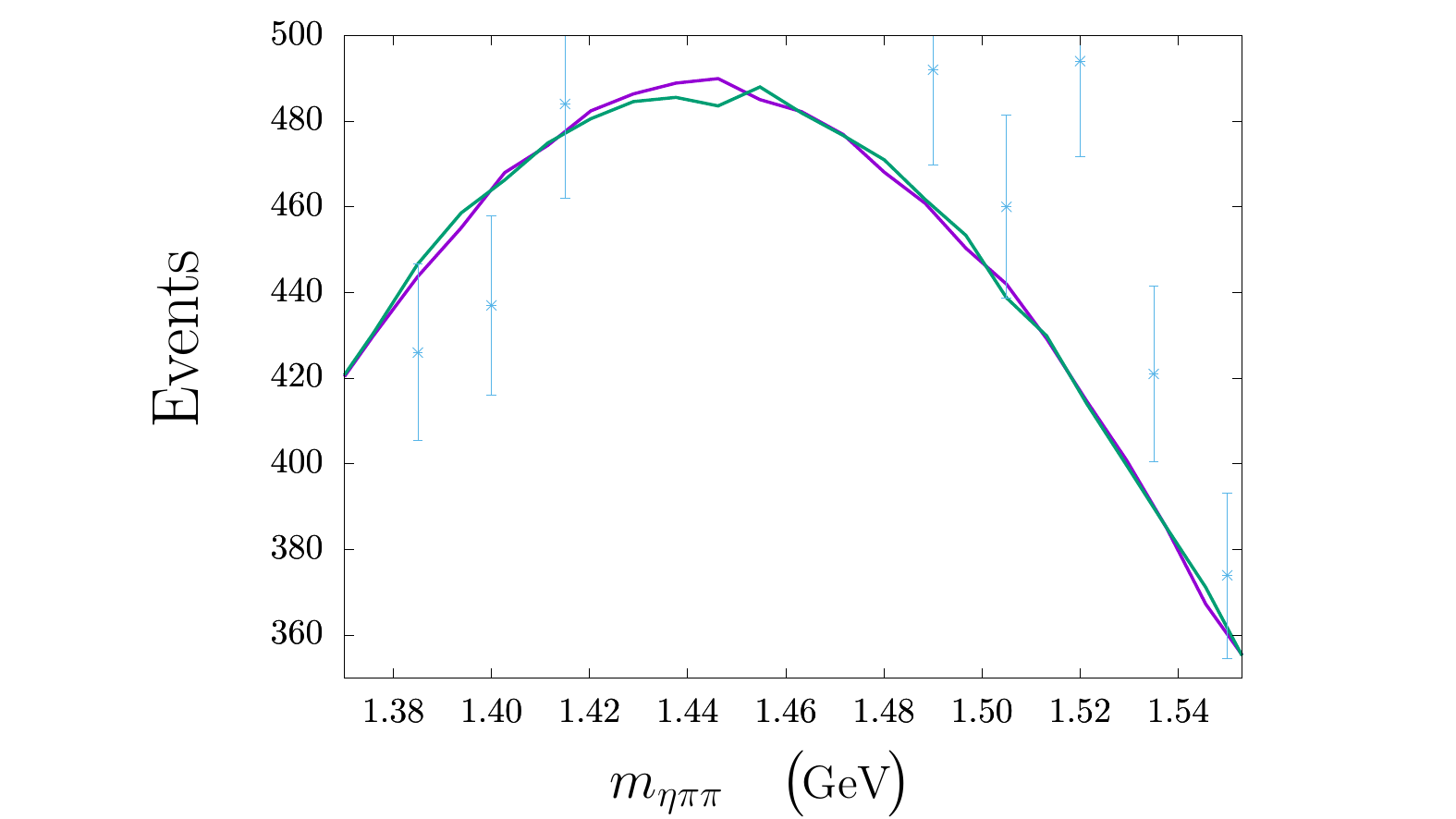}
\caption{Close up of Figure \ref{fig:SMvsNSI} in the small region where they differ.}\label{fig:CloseUp}
 \end{figure}

 When comparing the result of the total differential decay width to that obtained only from the SM contribution to the amplitude and the Belle spectrum, shown in Figure \ref{fig:SMvsNSI}, we confirm that the possible NSI contribution is undetectable with current data. In Fig. \ref{fig:CloseUp} we show a close up of the region where both curves of Fig. \ref{fig:SMvsNSI} differ a bit more.

 \begin{figure}[!ht]
  \centering\includegraphics[scale=0.85]{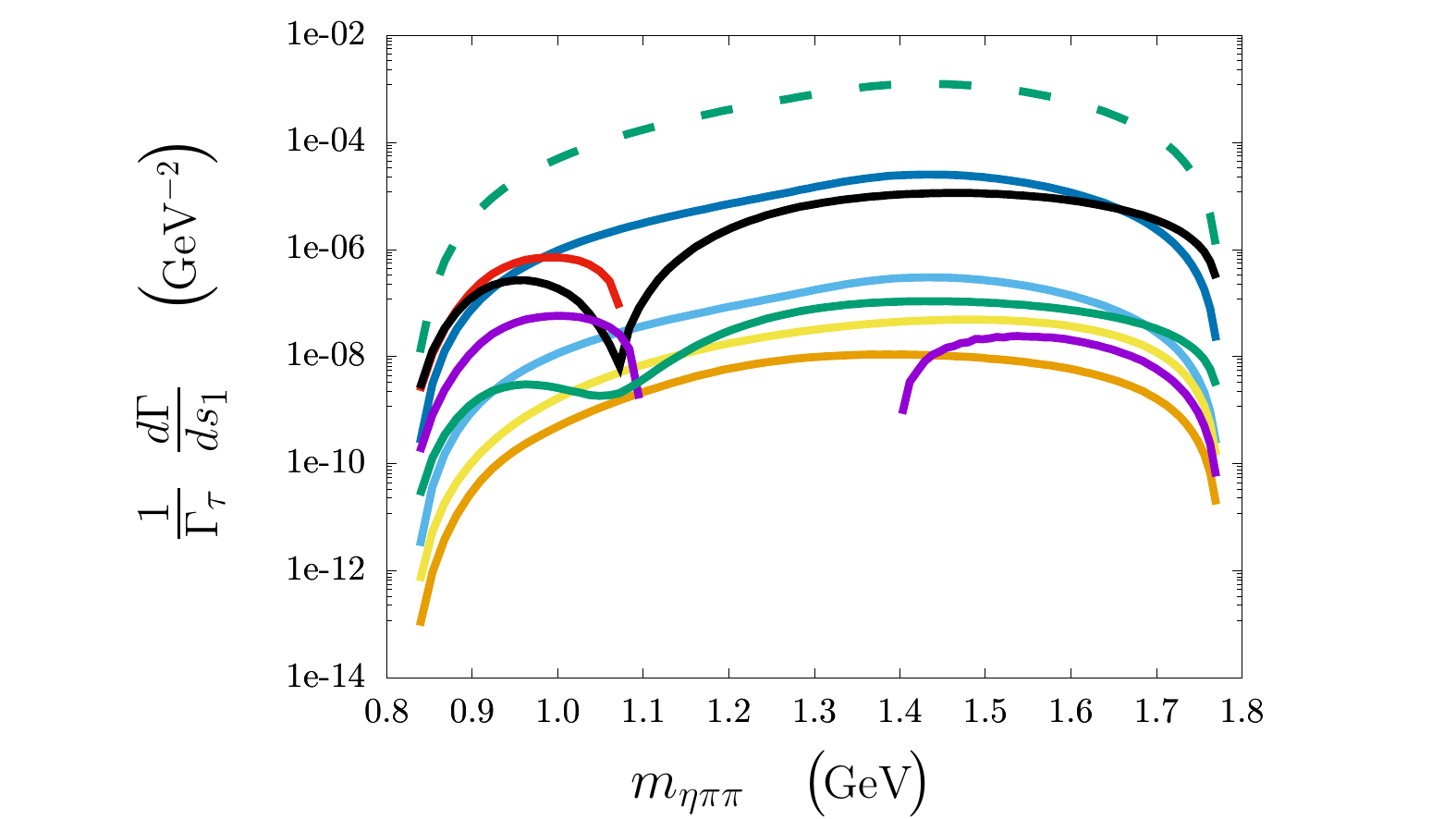}\caption{Different contributions to the differential decay width, shown in a logarithmic scale. Because of this, some of them cannot be shown in this figure, since interference terms are negative in certain regions of the invariant mass range. The color code follows: pure SM - dark green, SM-V - dark blue, SM-A - red, SM-T - black, A-V - lilac, V-T - green, A-T purple, $|$V$|^2$ - pale blue, $|$A$|^2$ - yellow, $|$T$|^2$ orange.}\label{fig:All}
 \end{figure}

\begin{figure}[!ht]
  \centering\includegraphics[scale=0.85]{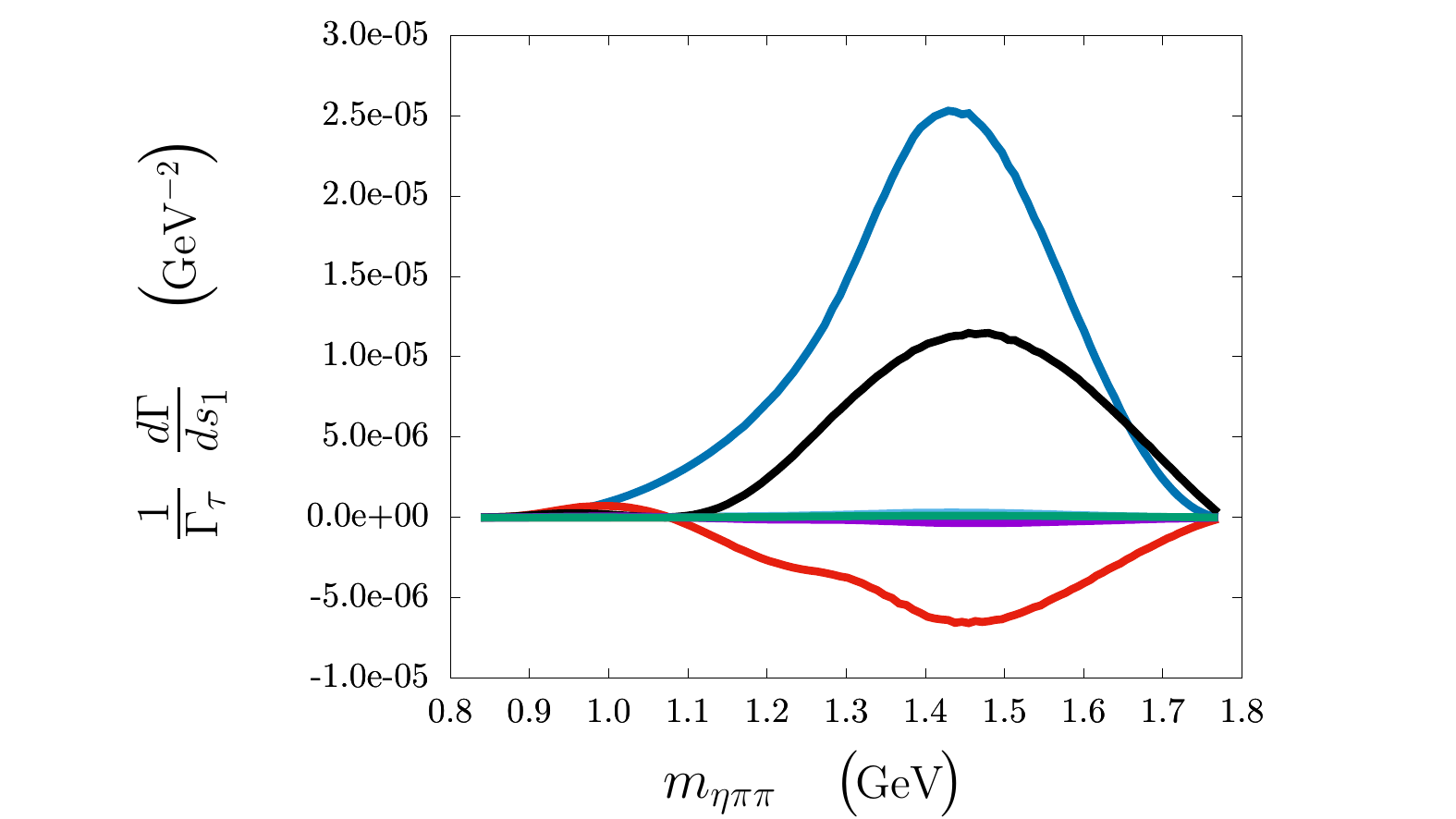}\caption{Same contributions from Figure \ref{fig:All}, without the SM contribution. The color code is the same as in the previous figure.}\label{fig:NoSM}
 \end{figure}

 We also obtained the contributions to the decay width from the different terms in the squared amplitude, this is, pure $V$, $A$, $P$, $T$, SM terms or only one of the interference terms among them, in turn. This is shown in Figure \ref{fig:All} in a logarithmic scale. In Fig. \ref{fig:NoSM} we show all such contributions, except for the pure SM one, in a normal scale. For most of the phase space the interference of the SM with the  vector non-SM interaction dominates. At low invariant masses there is a small window where the SM-tensor and SM-axial interferences  overcome it slightly. It is also seen  that the SM-tensor interference dominates near the endpoint.  It must be noted, however, that the tensor effects at high invariant masses may be smaller than depicted, as we are using for this form factor only the leading order contribution in the chiral expansion. Going beyond this approximation  should -in particular- reduce the effects shown for the SM-tensor interference at high $m_{\eta\pi\pi}$. %The clear dominance of the NSI vector effects agrees with the last uncertainty in eq. (\ref{eq:BRwithNSI}), which can be traced back to the bound on $\epsilon_V$ in eq.~(\ref{Ciriglianobounds2}). 
 These results can be used to study possible new physics effects in the $\tau^-\to\eta\pi^-\pi^0\nu_\tau$ decays with future improved data.

 \section{Conclusions}\label{sec:Concl}
 We have studied whether the discrepancy between isospin-rotated $\sigma(e^+e^-\to\eta\pi^+\pi^-)$ and $d\Gamma(\tau^-\to\eta\pi^+\pi^0\nu_\tau)/dm_{\eta\pi\pi}$ data can be explained by heavy new physics beyond the SM. Within an effective field theory approach for the NSI (assuming left-handed neutrinos), and using the bounds obtained previously on the corresponding new physics couplings,  we have shown that it is impossible to explain this tension between $e^+e^-$ and $\tau$ data by heavy new physics. Future measurement of the $\tau$ decay channel at Belle-II will shed light on the origin of this controversy. As a by-product of our analysis, we have developed the formalism needed to study NSI in three-meson tau decays (see ancillary file), which can be useful for other decay channels where hadronization is more complicated.

 \section*{Acknowledgements}
S.~A.~ acknowledges Conacyt for her Ms. Sc. scholarship at Cinvestav. L.-Y.~D.~ is  supported by Joint Large Scale Scientific Facility Funds of the National Natural Science Foundation of China (NSFC) and Chinese Academy of Sciences (CAS) under Contract No.U1932110, NSFC Grant with No.~12061141006, and Fundamental Research Funds for the Central Universities.
 P.~R.~ is indebted to Kenji Inami for providing him with Belle data  %\cite{Inami:2008ar} 
and for clarifying explanations on this analysis, and thanks Michel Hernández Villanueva and Iván Heredia de la Cruz work on this topic. Useful discussions on this subject with Gabriel López Castro and Antonio Rodríguez Sánchez are also acknowledged. P.~R.~ was partly funded by Conacyt’s project within ‘Paradigmas y Controversias de la Ciencia 2022’,
number 319395, and by Cátedra Marcos Moshinsky (Fundación Marcos Moshinsky) 2020, whose support is also acknowledged by A.~G.

\section*{Appendix: Brief overview of Resonance Chiral Theory}\label{sec:RChT}
In this appendix we recapitulate briefly the framework in which the model-dependent contributions have been evaluated \cite{Dai:2013joa,Qin:2020udp,Dumm:2012vb}, Resonance Chiral Theory (R$\chi$T) \cite{RChT}. See, for instance ref. \cite{Cirigliano:2006hb} for further details.

Resonances are added as explicit degrees of freedom to the $\chi$PT Lagrangian, which is enlarged by terms including them~\footnote{We note that $\chi$PT  operators coefficients are different in R$\chi$T according to the contributions, to the $\chi$PT low-energy constants, of integrating resonances out.}, where the $\chi$PT chiral tensors also appear. The symmetries determining the Lagrangian operators are the chiral one for the lightest pseudoscalar mesons (which are pseudoGoldstone bosons) and unitary symmetry for the resonances,  $SU(3)_L\otimes SU(3)_R\to SU(3)_V$ and $U(3)_V$, respectively, for the three lightest quark flavors. The expansion parameter of R$\chi$T is the inverse of the number of colors \cite{tHooft:1973alw}, where the leading order corresponds to tree level diagrams with an infinite tower of mesons per quantum number~\cite{tHooft:1973alw,tHooft:1974pnl} (the most important subleading correction comes from finite resonance widths).

Symmetries do not restrict the coupling values, so these should in principle be determined phenomenologically. However, assuming that the theory with resonances can interpolate between the chiral and parton regimes, Green functions in R$\chi$T need to comply with the known (from the corresponding operator product expansion) QCD short-distance behaviour. This determines or relates some of the couplings, increasing the predictivity of R$\chi$T. At the same time, this requirement tightly constrains contributions from operators with high-order chiral tensors. Complementary, the number of resonance fields is limited by the process at hand (via the number of initial and final state mesons, to which exchanged resonances couple). Altogether, this restricts, in practice, the number of operators of the R$\chi$T Lagrangians in the large-$N_C$ limit. 
The minimal interactions with (pseudo)scalar and (axial)vector resonances are given by \cite{RChT}
\begin{eqnarray}
\mathcal{L}_V&=&\frac{F_V}{2\sqrt{2}}\left\langle V_{\mu\nu}f_+^{\mu\nu}\right\rangle+i\frac{G_V}{\sqrt{2}}\left\langle V_{\mu\nu}u^{\mu}u^{\nu}\right\rangle\,,\;\mathcal{L}_A\,=\,\frac{F_A}{2\sqrt{2}}\left\langle A_{\mu\nu}f_-^{\mu\nu}\right\rangle\,,\nonumber\\
\mathcal{L}_S&=&c_d\left\langle S u^\mu u_\mu \right\rangle + c_m \left\langle S \chi_+ \right\rangle\,,\;\mathcal{L}_P\,=\, i d_m \left\langle P \chi_- \right\rangle\,,
\end{eqnarray}
see ref.~\cite{RChT} for further details.

Recent applications of R$\chi$T include  refs.~\cite{Guevara:2018rhj,Dai:2019lmj,Roig:2019reh,Miranda:2020wdg,Qin:2020udp,GutierrezSantiago:2020bhy,Arroyo-Urena:2021nil, Guevara:2021tpy, Arroyo-Urena:2021dfe, Chen:2022nxm}, mostly focused on tau decays and hadronic  contributions to the muon g-2.

\end{document}